\documentclass[prb,preprint]{revtex4}
\usepackage{amsfonts}
\usepackage[intlimits]{amsmath}
\usepackage{amssymb}
\usepackage[dvipdf]{graphicx}

\begin{document}

\title{A microscopic model for current-induced switching of magnetization for half-metallic leads}
\author{N. Sandschneider}
\email{niko.sandschneider@physik.hu-berlin.de}
\author{W. Nolting}
\affiliation{Institut f\"ur Physik, Humboldt-Universit\"at zu Berlin, Newtonstr. 15, 12489 Berlin, Germany}

\begin{abstract}
We study the behaviour of the magnetization in a half-metallic ferromagnet/nonmagnetic insulator/ferromagnetic metal/paramagnetic metal (FM1/NI/FM2/PM) tunnel junction. It is calculated self-consistently within the nonequilibrium Keldysh formalism. The magnetic regions are treated as band ferromagnets and are described by the single-band Hubbard model. We developed a nonequilibrium spectral density approach to solve the Hubbard model approximately in the switching magnet. By applying a voltage to the junction it is possible to switch between antiparallel (AP) and parallel (P) alignment of the magnetizations of the two ferromagnets. The transition from AP to P occurs for positive voltages while the inverse transition from P to AP can be induced by negative voltages only. This behaviour is in agreement with the Slonczewski model of current-induced switching and appears self-consistently within the model, i.e. without using half-classical methods like the Landau-Lifshitz-Gilbert equation.
\end{abstract}

\maketitle

There has been considerable interest in the phenomenon of current-induced switching of magnetization since it was first proposed over 10 years ago\cite{Slonczewski,Berger}. The basic idea behind this effect is as follows: the spin direction of electrons moving in a ferromagnet will be mostly aligned parallel to the magnetization axis. When these spin-polarized electrons are transported to a second ferromagnet, e.g. by applying a voltage, then the spin angular momentum of the itinerant electrons will exert a torque on the local magnetic moment. This torque is known as the spin-transfer torque. It will have an influence on the direction of magnetization. If the parameters of the materials are chosen in the right way and if the current through the junction is high enough it is even able to switch the magnetization of one ferromagnet from parallel to antiparallel or vice versa relative to the other one. This effect was seen both in all-metallic junctions\cite{Grollier,Myers,Tsoi} such as Co/Cu/Co and in magnetic tunnel junctions (MTJs) consisting of two ferromagnets divided by a thin nonmagnetic insulator\cite{Huai,Liu,Fuchs}. In this paper we focus on a special case of the latter, where the ferromagnetic lead is half-metallic, i.e. there are only electrons of one spin direction present at the Fermi energy.\\
Some of the possible technological applications of spin-transfer torques in MTJs have been discussed by Diao et al.\cite{Diao}. Most of the theoretical work in this area of research has been focused on the Landau-Lifshitz-Gilbert (LLG) equation\cite{Duine,Edwards,Sun,Weinberger,Zhang}, which is a macroscopic, half-classical equation. The torques entering this equation were usually calculated in a microscopic picture while treating the interactions on a mean field level. In this paper we propose a model which takes interactions beyond mean field into account. We make no use of the LLG equation or other macroscopic approaches and thus we stay on the quantum mechanical level throughout this paper.\\
\begin{figure}
\includegraphics[width=0.7\linewidth]{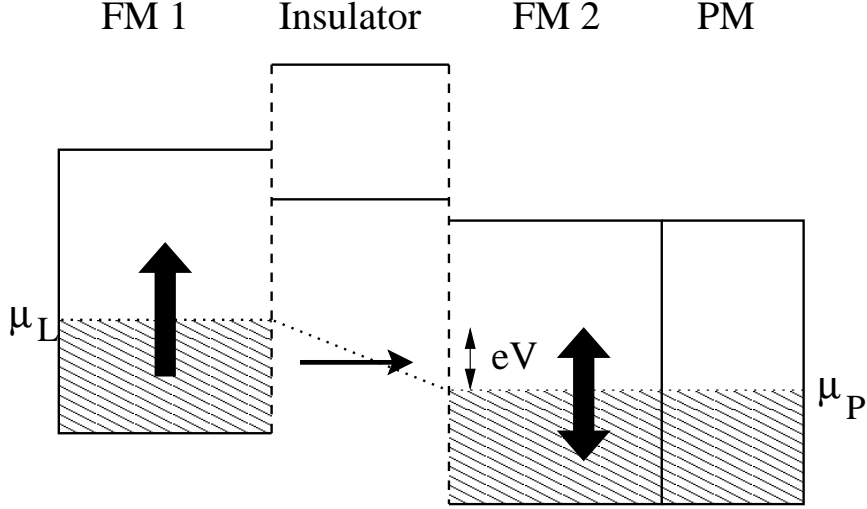}
\caption{Schematic picture of the magnetic tunnel junction with applied voltage V. The conduction bands are shown as rectangles. Occupied states in the metals are hatched and the directions of magnetization in the ferromagnets are symbolized by thick arrows.}\label{fig:MTJ}
\end{figure}
We will start the presentation of the theory by introducing a model Hamiltonian which describes the magnetic tunnel junction shown schematically in Fig. \ref{fig:MTJ}. There are two ferromagnetic metals ($L$ and $R$) divided by a non-magnetic insulator ($I$) and additionally a paramagnetic metal ($P$) which is necessary to have a well-defined chemical potential on the right side of the second ferromagnet. Each region consists of a single s-like band. The two outer leads $L$ and $P$ are treated as semi-infinite.\\
The total Hamiltonian consists of several parts:
\begin{equation}\label{eq:ModelH}
 H = H_L+H_{LI}+H_I+H_{RI}+H_R+H_{RP}+H_P
\end{equation}
$H_{L(R)}$ describes the left (right) ferromagnet, $H_I$ the insulator and $H_P$ the paramagnet. Both insulator and paramagnet are assumed to be non-interacting, so their Hamiltonians consist of the kinetic energy only:
\begin{equation}
 H_X = \sum_{\mathbf{k}_X\sigma} (\epsilon_{\mathbf{k}_X}-V_X) d_{\mathbf{k}_X\sigma}^+ d_{\mathbf{k}_X\sigma}\quad(X=I,P)
\end{equation}
$d_{\mathbf{k}_X\sigma}(d_{\mathbf{k}_X\sigma}^+)$ is the annihilation (creation) operator of an electron with wave vector $\mathbf{k}_X$ and spin $\sigma$. $\epsilon_{\mathbf{k}_X}$ is the dispersion of the lattice which throughout this paper is chosen as a tight-binding bcc lattice. The applied voltage $V$ will shift the center of gravity of the paramagnet by $V_P=V$ and half of that amount for the insulator, $V_I=V/2$. Positive voltage, $V>0$, will shift the bands to lower energies, while negative applied voltages result in a shift towards higher energies.\\
The Hamiltonians of the left (L) and right (R) ferromagnet are formally almost identical. Besides the kinetic energy they also include on-site Coulomb interaction. They are given in a mixed Bloch-Wannier representation:
\begin{equation}
 H_M = \sum_{\mathbf{k}_M\sigma} (\epsilon_{\mathbf{k}_M}-V_M) c_{\mathbf{k}_M\sigma}^+ c_{\mathbf{k}_M\sigma} + \frac{U_M}{2} \sum_{i_M\sigma} \hat{n}_{i_M\sigma}\hat{n}_{i_M-\sigma}\label{eq:H_M}
\end{equation}
where $M$ stands for either $L$ or $R$. The Hubbard-$U$ determines the interaction strength. $\hat{n}_{i_M\sigma}=c_{i_M\sigma}^+ c_{i_M\sigma}$ is the occupation number operator. The voltage $V$ shifts only the band center of the right ferromagnet, $V_R=V$, while the left ferromagnet is not directly influenced by $V$, i.e. $V_L=0$.\\
The remaining three terms of the Hamiltonian are responsible for the coupling between the different regions. These couplings act as a hybridization\cite{Sandschneider} between the bands and therefore the Hamiltonians are ($M=L,R;X=I,P$):
\begin{eqnarray}
 H_{MX} &=& \sum_{\mathbf{k}_M\mathbf{k}_X\sigma} \left(\epsilon_{\mathbf{k}_M\mathbf{k}_X} c_{\mathbf{k}_M\sigma}^+ d_{\mathbf{k}_X\sigma} + h.c.\right)\label{eq:H_MX}
\end{eqnarray}
They are characterized by the coupling constants $\epsilon_{\mathbf{k}_M\mathbf{k}_X}$ which determine the strength of the hybridization between the different bands. In general the couplings are wave vector dependent but for the sake of simplicity we neglect this dependence, $\epsilon_{\mathbf{k}_M\mathbf{k}_X}\equiv \epsilon_{MX}\equiv \epsilon_{XM}$. Furthermore we assume the coupling between the ferromagnets and the insulator to be equal, so that $\epsilon_{LI}=\epsilon_{RI}\equiv\epsilon_{MI}$. Altogether there remain two couplings $\epsilon_{MI}$ and $\epsilon_{RP}$ which cannot be calculated within this model, so they will be treated as parameters.\\
The main topic of this work will be the calculation of the non-equilibrium magnetization $m$ of the right ferromagnet within the Keldysh formalism\cite{Keldysh}. It can be calculated with the help of the Fourier transform of the so-called lesser Green's function defined as $G_{\mathbf{k}_R\sigma}^<(t,t') = i\langle c_{\mathbf{k}_R\sigma}^+(t') c_{\mathbf{k}_R\sigma}(t)\rangle$:
\begin{equation}
 m = n_\uparrow - n_\downarrow = \frac{1}{2\pi iN}\int_{-\infty}^{+\infty} dE \sum_{\mathbf{k}_R}\left( G_{\mathbf{k}_R\uparrow}^<(E)-G_{\mathbf{k}_R\downarrow}^<(E)\right)
\end{equation}
$n_\sigma=\langle \hat{n}_\sigma\rangle$ is the occupation number of particles with spin $\sigma$ in the right ferromagnet. In order to derive the lesser Green's function one first has to calculate the retarded one, $G_{\mathbf{k}_R\sigma}^r(E) = \langle\langle c_{\mathbf{k}_R\sigma};c_{\mathbf{k}_R\sigma}^+\rangle\rangle_E$. By using the equation of motion method one finds
\begin{equation}\label{eq:Gr}
G_{\mathbf{k}_R\sigma}^r(E) = \frac{1}{E-\epsilon_{\mathbf{k}_R}-\Sigma_{\mathbf{k}_R\sigma}^r(E)-\Delta_{\mathbf{k}_R\sigma}^r(E)}
\end{equation}
Two different selfenergies appear in the Green's function. First there is the interaction selfenergy which can only be calculated approximately for the Hubbard model. We propose a non-equilibrium spectral density approach (NSDA). The basic idea behind this approach is to choose the selfenergy in such a way that the first four spectral moments are reproduced by the theory. Some details of its derivation are given in Appendix \ref{app:NSDA}. The mean field (Stoner) solution of the Hubbard model on the other hand satisfies only the first two moments. One finds for the selfenergy
\begin{equation}\label{eq:selfenergy}
 \Sigma_{\mathbf{k_R}\sigma}^r(E) = U_R n_{-\sigma}\frac{E-T_{0,R}-B_{-\sigma}}{E-T_{0,R}-B_{-\sigma}-U_R(1-n_{-\sigma})}
\end{equation}
This expression is coincidentally formally identical to the equilibrium spectral density approach\cite{SDA}. The difference is in the spin-dependent band correction $B_{-\sigma}$ which is given by
\begin{widetext}
\begin{eqnarray}
 n_{-\sigma}(1-n_{-\sigma}) \bigl(&B_{-\sigma}&-T_{0,R}\bigr) = \frac{1}{2\pi iN} \sum_{\mathbf{k}_R}\int_{-\infty}^{\infty}dE\,\Bigl[\Bigl\{\left(\frac{2}{U_R}\Sigma_{\mathbf{k_R}-\sigma}^r(E)-1\right)
\left(E-T_{0,R}-\Sigma_{\mathbf{k_R}-\sigma}^r(E)\right) +\nonumber\\
 &+& \left(\frac{2}{U_R}-1\right)\left(\epsilon_{\mathbf{k}_R}-T_{0,R}\right)\left(E-\epsilon_{\mathbf{k}_R}-\Sigma_{\mathbf{k_R}-\sigma}^r(E)\right)\Bigr\}
G_{\mathbf{k}_R-\sigma}^<(E) + \frac{2}{U_R}\Delta_{\mathbf{k}_R-\sigma}^<(E)\Bigr]\label{eq:bandcorrection}
\end{eqnarray}
\end{widetext}
It has to be calculated self-consistently since $B_{-\sigma}$ also appears on the right hand side as part of the lesser Green's function. $T_{0,R}$ is the center of gravity of the right ferromagnet. The second selfenergy is the transport selfenergy which is due to electrons hopping between the different materials. Its retarded (lesser) component is given by
\begin{equation}\label{eq:Delta}
 \Delta_{\mathbf{k}_R\sigma}^{r(<)}(E) = \frac{1}{N}\sum_{\mathbf{k}_I}\epsilon_{MI}^2G_{\mathbf{k}_I\sigma}^{(L),r(<)}(E) +
 \frac{1}{N}\sum_{\mathbf{k}_P}\epsilon_{RP}^2g_{\mathbf{k}_P\sigma}^{r(<)}(E)
\end{equation}
$G_{\mathbf{k}_I\sigma}^{(L),r}(E)$ is the Green's function of the insulator when it is only coupled to the left ferromagnet, i.e.
\begin{equation}
 G_{\mathbf{k}_I\sigma}^{(L),r}(E) = \frac{1}{E-\epsilon_{\mathbf{k}_I}-\frac{1}{N}\sum_{\mathbf{k}_L}\epsilon_{MI}^2g_{\mathbf{k}_L\sigma}^r(E)}
\end{equation}
Since we neglected the wave vector dependence of the couplings, the transport selfenergy is only formally dependent on the wave vector. $g_{\mathbf{k}_L\sigma}^r(E)$ and $g_{\mathbf{k}_P\sigma}^r(E)$ are the equilibrium Green's functions of the left ferromagnet and the paramagnet, respectively. They can be easily calculated by the equation of motion method. One finds for $M=L,P$:
\begin{equation}
 g_{\mathbf{k}_M\sigma}^r(E) = \frac{1}{E-\epsilon_{\mathbf{k}_M}-\Sigma_{\mathbf{k}_M\sigma}^r(E)}
\end{equation}
The paramagnet does not include interactions so that $\Sigma_{\mathbf{k}_P\sigma}^r\equiv -i0^+$. Since we are mainly interested in the properties of the right ferromagnet, we assume that the left one is half-metallic so that its minority states play no role for small voltages. This is done by using the mean field selfenergy $\Sigma_{\mathbf{k}_L\sigma}^r=U_L n_{L,-\sigma}$ with sufficiently large $U_L$. Thus the retarded Green's function is known.\\
The lesser Green's function follows immediately from the Keldysh equation:
\begin{equation}
 G_{\mathbf{k}_R\sigma}^<(E) = G_{\mathbf{k}_R\sigma}^r(E)\Delta_{\mathbf{k}_R\sigma}^<(E) G_{\mathbf{k}_R\sigma}^a(E)
\end{equation}
where the advanced Green's function is simply the complex conjugated of the retarded one, $G_{\mathbf{k}_R\sigma}^a(E)=(G_{\mathbf{k}_R\sigma}^r(E))^*$. Furthermore we need the lesser component of the transport selfenergy which was already defined in Eq. (\ref{eq:Delta}). The lesser part of the insulator Green's function can again be calculated with the help of the Keldysh equation:
\begin{equation}
 G_{\mathbf{k}_I\sigma}^{(L),<}(E) = \frac{1}{N}\sum_{\mathbf{k}_L}G_{\mathbf{k}_I\sigma}^{(L),r}(E)\epsilon_{MI}^2g_{\mathbf{k}_L\sigma}^<(E)G_{\mathbf{k}_I\sigma}^{(L),a}(E)
\end{equation}
Since the Green's functions in the left ferromagnet and the paramagnet are equilibrium quantities, their lesser parts read
\begin{equation}
 g_{\mathbf{k}_{L(P)}\sigma}^<(E) = -2if_{L(P)}(E) \mbox{Im}\,g_{\mathbf{k}_{L(P)}\sigma}^r(E)
\end{equation}
where $f_{L(P)}(E)$ is the Fermi function in lead $L(P)$ with chemical potential $\mu_{L(P)}$. They are related by $\mu_L-\mu_P=V$. Thus we have a closed set of equations for calculating the magnetization of the ferromagnet.\\
\begin{figure}
\includegraphics[width=0.7\linewidth]{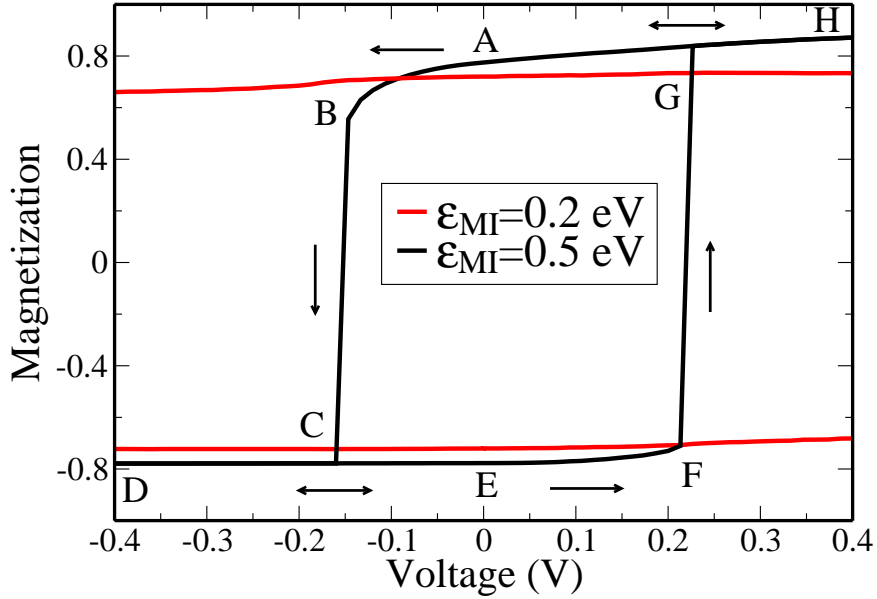}
\caption{Numerical results for the magnetization as a function of applied voltage for two different values of the hybridization strength $\epsilon_{MI}$ between the metals and the insulator. The arrows indicate in which direction the voltage was changed. Parameters: Band occupation $n=0.7$; band widths: $W_L=3\mbox{ eV},W_I=1\mbox{ eV},W_R=2$ eV and $W_P=5\mbox{ eV}$; interaction strengths: $U_R=4\mbox{ eV},U_L=4\mbox{ eV}$; band center of the insulator $T_{0,I}=5\mbox{ eV}$; hybridization strength: $\epsilon_{RP}=0.05$ eV; temperature: $T=0\mbox{ K}$}\label{fig:switch}
\end{figure}
In Fig. \ref{fig:switch} a typical numerical solution for the voltage-dependent magnetization is shown. We will first discuss the black curve, which was calculated with a hybridization strength of $\epsilon_{MI}=0.5$ eV. For the calculation we started with parallel alignment of the two magnetizations (point $A$ in the figure). Then a negative voltage is applied, i.e. the right ferromagnet is shifted to higher energies compared to the left one. At a critical voltage the parallel alignment becomes unstable and the magnetization reverses its sign ($B$ to $C$). Thus the magnetizations are now antiparallel. When the voltage is further decreased the magnetization stays more or less constant until point $D$ is reached. Then the process is reversed and the voltage is reduced to zero again. The magnetization follows the same line as before until the switching point $C$ is reached. There it does not switch back to parallel alignment but rather stays at about the same level. When $E$ is reached the direction of the voltage is reversed, i.e. the right ferromagnet will now be shifted to lower energies. For small voltages there is only a slight increase until a critical voltage is reached ($F$). This voltage has approximately the same value as the first one at point $B$, but of course with an opposite sign. There the antiparallel alignment is no longer stable and the systems returns to its initial parallel state which is not influenced by higher voltages ($G$ to $H$). Then the voltage is turned off and the system will be at its starting point $A$ again, so the hysteresis loop is complete.\\
As another test we start again at point $A$, but this time we turn on a positive voltage. Then no switching occurs and the system will move reversibly to point $H$. A similar reversible behaviour is seen when the alignment is antiparallel ($E$) and the voltage is decreased. This is shown by the arrows in the figure. So, one has to conclude that switching of magnetization from parallel to antiparallel alignment is only possible for negative voltages and the reverse process will only appear for positive voltage. The behaviour just described is one of the hallmarks of current-induced switching of magnetization and thus our proposed model is indeed able to simulate this effect without leaving the microscopic picture!\\
\begin{figure}
 \includegraphics[width=0.7\linewidth]{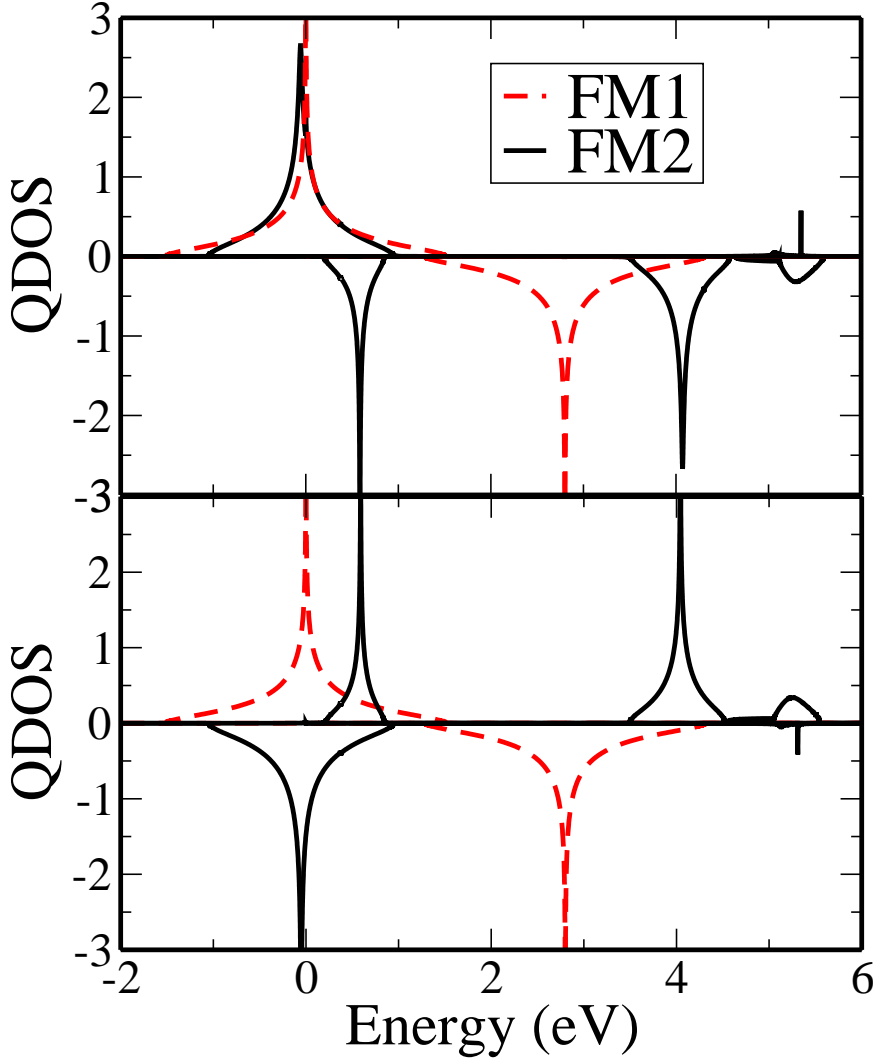}
 \caption{(Color online) Quasiparticle density of states for the parameter set of Fig. \ref{fig:switch} for $V=0$ and $\epsilon_{MI}=0.5$ eV. Upper picture for parallel alignment of the magnetizations (point A), lower picture for antiparallel alignment (point E). Spin up is shown along the positive, spin down along the negative axis. The black line is the QDOS of the right ferromagnet, treated within the NSDA, and the red broken line shows the QDOS of the left ferromagnet in mean field.}\label{fig:qdos}
\end{figure}
Now we want to give a short explanation on how exactly our model is able to provide these results. The key to the understanding lies in the effect the hybridization parts of the Hamiltonian have on the quasiparticle density of states (QDOS) of the switching magnet and in the polarization of the current. A hybridization between two bands generally will lead to a repulsion between them, i.e. the energetic distance between their respective centers of gravity will increase the stronger the effect of the hybridization is. The magnitude of this shift is mainly influenced by three quantities: the strength of the hybridization itself ($\epsilon_{RP}$ and $\epsilon_{MI}$ in our case), the energetic distance between the two bands (the closer they are to each other, the stronger they will be repelled) and their spectral weight (higher spectral weight leads to stronger repulsion). In the upper part of Fig. \ref{fig:qdos} we plotted a typical QDOS for the NSDA without applied voltage for parallel alignment of the two magnetizations. The dashed line represents the density of states of the left ferromagnet. The splitting of the right QDOS into lower and upper Hubbard band at $E\approx 0$ and $E\approx U_R$ is clearly visible. Additionally there are contributions of the insulator at $E\approx T_{0,I}=5$ eV which are due to the hybridization. What happens when a voltage is turned on depends on its sign. The left ferromagnet FM1 is not influenced by the voltage, so its QDOS will be the same. Let us first discuss the case $V>0$ where both spin bands of the right ferromagnet FM2 are shifted to lower energies. But due to the repulsion between the spin up bands of FM1 and FM2 the shift of the spin up band will actually be stronger than for spin down, where the repulsion is much weaker. So a positive voltage leads to a stabilization of the magnetization. This effect is enhanced by the current. For positive voltages it will flow from left to right. Since the left ferromagnet is fully polarized there are only spin up electrons tunneling into the right ferromagnet. These are the reasons for the slight increase of magnetization in Fig. \ref{fig:switch} between points A and H and also an explanation why there can be no switching in this case. On the other hand, if we apply a negative voltage, $V<0$, the right spin bands will be shifted to higher energies. For the same reasons as discussed above the shift of the spin up band will be enhanced by the hydridization. Thus the difference between the centers of gravity of both spin bands will be decreased. If the hybridization strength is sufficiently large this additional shift together with the selfconsistency will be enough to push the spin up band above the spin down band and thus the magnetization changes sign. The selfconsistency is important since it will enhance the shift, because the occupation in one band depends on the occupation in the other one. In this case current flows from the right to the left lead. Since there are only spin up states available for tunneling in the left ferromagnet, the tunneling current flowing out of the switching magnet will consist of spin up electrons only. This leads to an additional decrease of the right magnetization. This explains the behaviour shown by the parallel aligned curve in Fig. \ref{fig:switch}.\\
The antiparallel case can be explained in a very similar way. In the lower part of Fig. \ref{fig:qdos} the corresponding quasiparticle density of states is shown, again without applied voltage. The obvious difference to the case discussed above is that the center of gravity of the lower spin down band is below the lower spin up band. This is the reason for the negative magnetization of course, but it is also responsible for the reversed behaviour with respect to the applied voltage. In this case a negative voltage cannot push the center of gravity of the spin up band below the spin down band. It rather has the opposite effect because the repulsion pushes the spin up band to even higher energies and thus leads to a more stable magnetization which can also be seen in Fig. \ref{fig:switch} between points E and D. For a positive voltage the spin up band of FM2 moves below the spin up band of FM1 so that the hybridization will shift it to lower energies compared to the spin down band. Again, if the hybridization strength is larger than a critical value this additional shift will be enough to reverse the two spin directions such that the magnetization changes sign. For the same reasons as discussed for the parallel case, a positive voltage will increase the magnetization, while a negative voltage has the opposite effect. Therefore the behaviour of the magnetization in Fig. \ref{fig:switch} can be understood in terms of the quasiparticle density of states.\\
In order to prove the explanation based on the hybridization we plotted a second magnetization curve in Fig. \ref{fig:switch} with smaller hybridization strength $\epsilon_{MI}=0.2$ eV. Obviously in this case no switching occurs. There is only a slight change of magnetization. Starting from parallel (antiparallel) alignment the magnetization is reduced for negative (positive) voltages. This is in agreement with the explanation given above. Since the hybridization is weaker the repulsion between the bands is also reduced. It is not strong enough to push the spin down band above the spin up band or vice versa. Thus the direction of magnetization is not changed. The current density through the junction is closely linked to the coupling strength between the materials\cite{Sandschneider}: smaller $\epsilon_{MI}$ corresponds to a weaker current. From the results shown in Fig. \ref{fig:switch} we can conclude that in order to switch the magnetization the current has to exceed a certain value.\\
To summarize, we presented a selfconsistent calculation of the voltage-dependent magnetization in a magnetic tunnel junction within a microscopic non-equilibrium framework. The magnetization shows a hysteresis behaviour similar to that seen in experiments. The reason for this effect was explained to be the hybridization between left and right ferromagnets which could be seen with the help of the quasiparticle density of states. It should be noted that the behaviour discussed above does only appear for very special parameter sets (such as low band occupation, small U) when one uses the mean field approximation for the right ferromagnet. This seems reasonable because it is known that mean field strongly overestimates the stability of ferromagnetism. Thus it should be more difficult to switch the direction of magnetization. We have to conclude that higher correlations seem to be an important factor when describing current-induced switching of magnetization within this model. One might argue that the Kondo peak is missing in the NSDA which should have considerable influence on the magnetization. However, we investigated the strong-coupling regime only, where it is known that the Kondo peak does not play a major role. On the other hand it would be a very interesting expansion of the model to examine its weak-coupling behaviour. Another important extension would be the inclusion of spin-orbit coupling which is widely believed to be the microscopic origin of phenomenological damping effects\cite{Hickey} which play a crucial role in the macroscopic description of switching of magnetization.

\begin{appendix}
\section{Non-equilibrium spectral density approach}\label{app:NSDA}
 The basic idea behind the nonequilibrium spectral density approach (NSDA) is to choose the selfenergy in such a way, that the first four spectral moments
\begin{equation}
 M_{\mathbf{k}_R\sigma}^{(n)} = -\frac{1}{\pi}\int_{-\infty}^\infty dE\,E^n \mbox{Im}\,G_{\mathbf{k}_R\sigma}^r(E)
\end{equation}
 of the spectral density are reproduced exactly. The moments are calculated with the help of the following exact relation:
\begin{equation}
 M_{\mathbf{k}_R\sigma}^{(n)} = \frac{1}{N}\sum_{i_Rj_R}e^{-\mathbf{k}_R\cdot(\mathbf{R}_{i_R}-\mathbf{R}_{j_R})}\left\langle[[\dots[c_{i_R\sigma},H]_-\dots,H]_-,[H\dots[H,c_{j_R\sigma}^+]_-\dots]_-]_+\right\rangle
\end{equation}
where the total number of commutators on the right hand side must be equal to $n$. Inserting the Hamiltonian (\ref{eq:ModelH}) into this expression yields after some calculation:
\begin{eqnarray}
 M_{\mathbf{k}_R\sigma}^{(0)} &=& 1\\
 M_{\mathbf{k}_R\sigma}^{(1)} &=& \epsilon_{\mathbf{k}_R} + U_Rn_{-\sigma}\\
 M_{\mathbf{k}_R\sigma}^{(2)} &=& \epsilon_{\mathbf{k}_R}^2 + 2U_R\epsilon_{\mathbf{k}_R}n_{-\sigma} + U_R^2 n_{-\sigma} + \epsilon_{MI}^2 + \epsilon_{RP}^2\\
 M_{\mathbf{k}_R\sigma}^{(3)} &=& \epsilon_{\mathbf{k}_R}^3 + 2\epsilon_{\mathbf{k}_R}(\epsilon_{MI}^2+\epsilon_{RP}^2)+\epsilon_{MI}^2T_{0,I} + \epsilon_{RP}^2T_{0,P} + U_R\Bigl\{3\epsilon_{\mathbf{k}_R}^2 n_{-\sigma} + 2(\epsilon_{MI}^2+\epsilon_{RP}^2) n_{-\sigma}\Bigr\} +\nonumber\\
& & + U_R^2\Bigl\{(2+n_{-\sigma})\epsilon_{\mathbf{k}_R}n_{-\sigma} + n_{-\sigma}(1-n_{-\sigma})B_{-\sigma}\Bigr\} + U_R^3n_{-\sigma}
\end{eqnarray}
The moments of the transport selfenergy $\Delta_{\mathbf{k}_R\sigma}^r(E)$ can be derived in the same way. One gets:
\begin{eqnarray}
 D_{\mathbf{k}_R\sigma}^{(0)} &=& 0\\
 D_{\mathbf{k}_R\sigma}^{(1)} &=& \epsilon_{MI}^2+\epsilon_{RP}^2\\
 D_{\mathbf{k}_R\sigma}^{(2)} &=& \epsilon_{MI}^2T_{0,I} + \epsilon_{RP}^2T_{0,P}
\end{eqnarray}
The band correction $B_{-\sigma}$ is given by
\begin{eqnarray}
 n_{-\sigma}(1-n_{-\sigma})(B_{-\sigma}-T_{0,R}) &=& \frac{1}{N}\sum_{i_Rj_R}(T_{i_Rj_R}-T_{0,R})\langle c_{i_R-\sigma}^+c_{j_R-\sigma}(2\hat{n}_{i_R\sigma}-1)\rangle +\nonumber\\
& &+ \frac{1}{N}\sum_{X=I,P}\sum_{i_Ri_X}(T_{i_Xi_R}\langle d_{i_X-\sigma}^+c_{i_R-\sigma}(2\hat{n}_{i_R\sigma}-1)\rangle
\end{eqnarray}
where $T_{i_Rj_R}$ is the hopping integral between lattice sites $\mathbf{R}_{i_R}$ and $\mathbf{R}_{j_R}$. The two higher correlation functions can be reduced to single-particle lesser Green's functions\cite{SDA}. We find
\begin{eqnarray}
 \langle d_{i_X-\sigma}^+c_{i_R-\sigma}\hat{n}_{i_R\sigma}\rangle &=& \frac{i}{2\pi NU_R}\sum_{\mathbf{k}_R\mathbf{k}_X}\int_{-\infty}^\infty dE\,e^{i(\mathbf{k}_R\cdot \mathbf{R}_{i_R}-\mathbf{k}_X\cdot\mathbf{R}_{i_X})}\Bigl[(-E+\epsilon_{\mathbf{k}_R}+\Delta_{\mathbf{k}_R-\sigma}^r(E))\cdot\nonumber\\
& &\cdot G_{\mathbf{k}_R\mathbf{k}_X-\sigma}^<(E) + \Delta_{\mathbf{k}_R-\sigma}^<(E)G_{\mathbf{k}_R\mathbf{k}_X-\sigma}^a(E)\Bigr]
\end{eqnarray}
and
\begin{equation}
 \langle c_{i_R-\sigma}^+c_{j_R-\sigma}\hat{n}_{i_R\sigma}\rangle = -\frac{i}{2\pi NU_R}\sum_{\mathbf{k}_R}e^{i\mathbf{k}_R\cdot(\mathbf{R}_{j_R}-\mathbf{R}_{i_R})}\int_{-\infty}^{\infty}dE\,\Sigma_{\mathbf{k}_R-\sigma}^r(E)G_{\mathbf{k}_R-\sigma}^<(E).
\end{equation}
The non-diagonal lesser Green's function $G_{\mathbf{k}_R\mathbf{k}_X\sigma}^<(E)= i\langle d_{\mathbf{k}_X\sigma}^+c_{\mathbf{k}_X\sigma}\rangle$ is closely related to the right Green's function and the transport selfenergy:
\begin{equation}
 \sum_{X=I,P}\sum_{\mathbf{k}_X}G_{\mathbf{k}_R\mathbf{k}_X\sigma}^<\epsilon_{XR} = G_{\mathbf{k}_R\sigma}^r\Delta_{\mathbf{k}_R\sigma}^< + G_{\mathbf{k}_R\sigma}^<\Delta_{\mathbf{k}_R\sigma}^a
\end{equation}
Putting all these expressions into the band correction leads to the result in Eq. \ref{eq:bandcorrection}.\\
The Dyson equation of the right ferromagnet reads
\begin{equation}
 E G_{\mathbf{k}_R\sigma}^r(E) = 1 + (\epsilon_{\mathbf{k}_R}+\Sigma_{\mathbf{k}_R\sigma}^r(E)+\Delta_{\mathbf{k}_R\sigma}^r(E))G_{\mathbf{k}_R\sigma}^r(E)
\end{equation}
Inserting the high-energy expansion for both selfenergies and the Green's function
\begin{eqnarray}
 G_{\mathbf{k}_R\sigma}^r(E) &=& \sum_{n=0}^\infty\frac{M_{\mathbf{k}_R\sigma}^{(n)}}{E^{n+1}}\\
 \Delta_{\mathbf{k}_R\sigma}^r(E) &=& \sum_{m=0}^\infty\frac{D_{\mathbf{k}_R\sigma}^{(m)}}{E^m}\\
 \Sigma_{\mathbf{k}_R\sigma}^r(E) &=& \sum_{m=0}^\infty\frac{C_{\mathbf{k}_R\sigma}^{(m)}}{E^m}\label{eq:Sigma_exp}
\end{eqnarray}
yields a system of equations for the unknown moments $C_{\mathbf{k}_R\sigma}^{(m)}$ of the interaction selfenergy. It can be solved by sorting according to the order of $1/E$ and the use of the moments given earlier in this appendix. The results are quite simple:
\begin{eqnarray}
 C_{\mathbf{k}_R\sigma}^{(0)} &=& U_Rn_{-\sigma}\\
 C_{\mathbf{k}_R\sigma}^{(1)} &=& U_R^2n_{-\sigma}(1-n_{-\sigma})\\
 C_{\mathbf{k}_R\sigma}^{(2)} &=& U_R^2n_{-\sigma}(1-n_{-\sigma})B_{-\sigma} + U_R^3n_{-\sigma}(1-n_{-\sigma})^2
\end{eqnarray}
These expressions are formally identical to the equilibrium case, therefore the selfenergy will also have the same form Eq. (\ref{eq:selfenergy})\cite{SDA}. For high energies it is acceptable to neglect higher order terms of the expansion in Eq. (\ref{eq:Sigma_exp}). Thus:
\begin{eqnarray}
 \Sigma_{\mathbf{k}_R\sigma}(E) &\approx& C_{\mathbf{k}_R\sigma}^{(0)}+\frac{C_{\mathbf{k}_R\sigma}^{(1)}}{E}+\frac{C_{\mathbf{k}_R\sigma}^{(2)}}{E^2}\nonumber\\
&\approx& C_{\mathbf{k}_R\sigma}^{(0)} + \frac{C_{\mathbf{k}_R\sigma}^{(1)}}{E-\frac{C_{\mathbf{k}_R\sigma}^{(2)}}{C_{\mathbf{k}_R\sigma}^{(1)}}}\nonumber\\
&=& U_R n_{-\sigma}\frac{E-T_{0,R}-B_{-\sigma}}{E-T_{0,R}-B_{-\sigma}-U_R(1-n_{-\sigma})}
\end{eqnarray}

\end{appendix}

\end{document}